\PassOptionsToPackage{unicode}{hyperref}
\PassOptionsToPackage{hyphens}{url}
\documentclass[
]{article}
\usepackage{xcolor}
\usepackage{amsmath,amssymb}
\usepackage{iftex}
\ifPDFTeX
  \usepackage[T1]{fontenc}
  \usepackage[utf8]{inputenc}
  \usepackage{textcomp} 
\else 
  \usepackage{unicode-math} 
  \defaultfontfeatures{Scale=MatchLowercase}
  \defaultfontfeatures[\rmfamily]{Ligatures=TeX,Scale=1}
\fi
\usepackage{lmodern}
\ifPDFTeX\else
\fi
\IfFileExists{upquote.sty}{\usepackage{upquote}}{}
\IfFileExists{microtype.sty}{
  \usepackage[]{microtype}
  \UseMicrotypeSet[protrusion]{basicmath} 
}{}
\makeatletter
\@ifundefined{KOMAClassName}{
  \IfFileExists{parskip.sty}{%
    \usepackage{parskip}
  }{
    \setlength{\parindent}{0pt}
    \setlength{\parskip}{6pt plus 2pt minus 1pt}}
}{
  \KOMAoptions{parskip=half}}
\makeatother
\usepackage{bookmark}
\IfFileExists{xurl.sty}{\usepackage{xurl}}{} 
\hypersetup{
  hidelinks,
  pdfcreator={LaTeX via pandoc}}

\title{A MATLAB Toolbox for Standardized Reading Speed Assessment:\\
Implementing and Extending the Perrin Sentence Generator for English Corpora}
\author{Daniel P. Spiegel\textsuperscript{1} \and Romain Bachy\textsuperscript{1}\\[0.5em]
\textsuperscript{1}Meta Reality Labs, 9805 Willows Rd, Redmond, WA 98052, USA}
\date{}

\begin{document}

\maketitle

\begin{abstract}
In the fields of vision science, cognitive psychology, and psycholinguistics, the accurate measurement of reading speed is frequently hampered by the limitations of static reading charts. Repeated testing often leads to memorization effects, while the requirement for oral recitation introduces speech-motor confounds that obscure true information processing speed. To address these methodological hurdles, this paper introduces an open-source MATLAB toolbox that adapts the sentence generation paradigm originally proposed by Perrin, Paillé, and Baccino (2014) for the English language. This system utilizes a semantic ontology and a ``proto-truth'' logic to autonomously generate thousands of unique, grammatically simple sentences with unambiguous truth values. Beyond the original scope of Maximum Reading Speed (MRS) measurement, this implementation introduces band-pass psycholinguistic filtering and specific logic to resolve semantic ambiguities unique to English. We present this complete software package as an open platform for the scientific community to validate and refine.
\end{abstract}

\section*{1. Introduction}

The assessment of reading speed is a cornerstone of clinical ophthalmology and experimental psychology, serving as a vital metric for evaluating visual function and the impact of low-vision pathologies (Ahn, Legge, \& Luebker, 1995; Seiple et al., 2005). However, the "gold standard" instruments traditionally used for this task, such as the MNREAD charts (Legge, Ross, Luebker, \& LaMay, 1989), face inherent limitations when longitudinal monitoring is required. The finite number of sentences on a printed chart means that patients undergoing repeated testing may inadvertently memorize the content, artificially inflating their scores. Furthermore, standard protocols often rely on oral reading, a task that conflates visual information acquisition with the motor planning required for speech.

To overcome these constraints, Crossland, Legge, and Dakin (2008) developed an automated sentence generator to provide thousands of sentences intended for reading speed assessment. Their algorithm randomly combined a quantifier, an object, and a trait to form grammatically valid sentences (e.g., "No chimps have feathers"). While a significant advancement for repeated measurements, this early generator occasionally produced ambiguous texts that caused reader comprehension errors independent of their actual visual processing speed.

To address this ambiguity, Perrin, Paillé, and Baccino (2014) introduced an upgraded computational approach. Their method utilizes a semantic ontology to structure concepts, allowing the system to automatically assign an unambiguous "True" or "False" value to every generated sentence without manual intervention. This ensures sentences are logically determining yet semantically simple, permitting a "silent reading" paradigm where effective reading is verified not by oral recitation, but by a simple True/False decision. This paper details the architecture of a MATLAB-based implementation of the Perrin generator, specifically adapted for English corpora through the integration of the Glasgow Norms (Scott et al., 2019).

\section*{2. The Logic of Automatic Sentence Generation}

The core challenge in automating reading materials is ensuring that the computer "knows" whether a generated sentence is true or false without human intervention. The Perrin method solves this by structuring the world not as a dictionary, but as a rooted tree ontology.

\subsection*{2.1 The Ontology as a Knowledge Map}

At the heart of the generator lies a hierarchical dataset representing concepts as nodes in a tree O=(N,E,r). These nodes are divided into "Classes" (broad categories like animal, fruit, or tool) and "Entities" (specific items like dog, apple, or hammer). The system determines the relationship between any two concepts based solely on their relative positions in the tree. For example, because the node for dog is a descendant of the node for mammal, which is in turn a descendant of animal, the system possesses an intrinsic "understanding" of these taxonomic relationships.

\subsection*{2.2 From Structure to ``Proto-Truth''}

When the generator selects two concepts to form a sentence O=(\emph{e\textsubscript{1}}, \emph{e\textsubscript{2}}), it first calculates a "proto-truth" value (\emph{pt}) based on their ancestry:

\begin{itemize}
\item
  \textbf{True (T):} If \emph{e\textsubscript{1}} is a descendant of \emph{e\textsubscript{2}} (e.g., "A dog is an animal").
\item
  \textbf{Contingent (T/F):} If the relationship is inverted (e.g., "An animal is a dog").
\item
  \textbf{False (F):} If the nodes share no direct lineage (e.g., "A dog is a fruit").
\end{itemize}

\subsection*{2.3 The Role of Quantifiers and Truth Tables}

To convert these abstract relationships into natural language, the system applies one of three quantifiers: "All," "Some," or "No". A truth-value mapping function then determines the final veracity of the statement. For instance, applying the quantifier "No" to a Proto-False pair (e.g., "A dog is not a fruit") results in a True statement.

Crucially, the system excludes specific combinations to prevent pragmatic ambiguity. The construction "Some + Proto-True" (e.g., "Some dogs are animals") is strictly forbidden; while logically true, it carries a Gricean implicature that some dogs are not animals, potentially confusing the reader (Perrin et al., 2014). The remaining valid combinations are weighted to ensure that the probability of a "True" sentence remains exactly 0.5.

\section*{3. Psycholinguistic Constraints}

To ensure homogeneity across thousands of generated sentences, this implementation filters the English lexicon using the Glasgow Norms dataset of 5,554 words (Scott et al., 2019). The Glasgow Norms are available under a Creative Commons Attribution 4.0 International License (CC BY 4.0); the original dataset is accessible at https://doi.org/10.3758/s13428-018-1099-3.

\subsection*{3.1 Familiarity, Concreteness, and Valence}

The toolbox selects words based on three standardized dimensions:

\begin{enumerate}
\def\labelenumi{\arabic{enumi}.}
\item
  Familiarity (FAM): To ensure reading speed reflects visual processing rather than vocabulary retrieval, the default setting requires FAM≥4.5 on a 7-point scale. Familiarity is strongly correlated with objective lexical frequency (Gernsbacher, 1984; Gilhooly \& Logie, 1980).
\item
  Concreteness (CNC): Based on Dual Coding Theory (Paivio, 1971; Paivio, Yuille, \& Madigan, 1968), concrete words activate both verbal and imaginal systems and are processed faster than abstract ones. The default filter requires CNC≥4.5.
\item
  Valence (VAL): To avoid the "negativity bias" where negative emotional content captures attention and slows processing (Baumeister et al., 2001; Kensinger \& Corkin, 2004), the system requires VAL ≥ 4.0, restricting the corpus to neutral and positive concepts (Frijda, 1986; Russell, 1980).
\end{enumerate}

\subsection*{3.2 Homonym Exclusion}

Words with multiple distinct meanings (e.g., bank, arm) are explicitly filtered out to avoid polysemous ambiguity (Perrin et al., 2014). Any entry in the norms dataset containing parenthetical disambiguation is removed from the candidate pool.

\section*{4. Methods and Implementation}

Adapting the French method to English required specific structural adjustments to accommodate grammatical constraints and semantic ambiguities unique to English morphology. The current ontology relies on 97 entities and 23 classes.

\subsection*{4.1 Ontology Construction and Mass Noun Filtration}

The sentence templates utilize pluralization for both subject and predicate (e.g., all \emph{e\textsubscript{1}} are \emph{e\textsubscript{2}}). This structure necessitates that all concepts be count nouns. Mass nouns (uncountable nouns) were systematically identified and removed to prevent ungrammatical generation.

\begin{itemize}
\item
  \textbf{Entity Removal:} Eleven entities were removed from the original corpus because they lacked grammatical plural forms suitable for the templates. These include meats (\emph{beef, pork, bacon, ham}), grains (\emph{rice, wheat, oats}), and other substances (\emph{milk, butter, honey, corn}). Additionally, \emph{scissors} was removed as it is a \emph{plurale tantum} noun with no singular form.
\item
  \textbf{Class Renaming:} Top-level classes that functioned as mass nouns were renamed to count noun equivalents. The class clothing was renamed to garment (to avoid \emph{"All shirts are clothings"}), and furniture was renamed to furnishing.
\item
  \textbf{Subclass Restructuring:} The subclasses dairy and grain were dissolved. Their remaining valid count-noun entities (e.g., \emph{cheese, yogurt, bread}) were moved directly under the parent food class.
\end{itemize}

\subsection*{4.2 Grammatical Standardization}

To ensure robust pluralization, the toolbox utilizes a lookup table for irregular nouns, expanded to 38 entries. This addresses invariant plurals (e.g., aircraft, watercraft) and specific English orthographic rules that standard regular expressions miss (e.g., bus → buses, oats → oats). Although removed from the active generation pool, mass nouns like beef are retained in the irregulars table as invariant forms to prevent runtime errors if users manually re-introduce them.

\subsection*{4.3 Cross-Branch Semantic Ambiguity Filtering}

A distinct logical problem arises in the "False" condition when ontological truth contradicts real-world knowledge. In the ontology, animal and food are disjoint branches; therefore, the generator calculates a "False" proto-truth value for pairs crossing these branches. However, in the real world, many animals are edible.

To prevent the generation of sentences that are ontologically "False" but empirically "True" (e.g., "Some frogs are foods" or "A chicken is not a food"), the system implements an isAmbiguousCrossBranch filter. This function specifically targets pairs where one element belongs to the animal branch and the other to the food branch. If the proto-truth value is False, the pair is discarded. This ensures that "False" sentences rely on genuine semantic disjointness (e.g., "A dog is not a vehicle") rather than biological technicalities.

\subsection*{4.4 Parametric Difficulty Control: Band-Pass Filtering}

To extend the utility of the toolbox beyond Maximum Reading Speed (MRS) measurement, the implementation replaces simple minimum thresholds with "Band-Pass" filtering. While the original method used only floor thresholds (e.g., FAM≥4.5), this approach allowed easy words to dominate the distribution even when the floor was lowered. The updated method introduces ceiling parameters (MaxFamiliarity, MaxConcreteness, MaxValence) to isolate specific difficulty ranges.

The toolbox includes three standardized presets:

\begin{enumerate}
\def\labelenumi{\arabic{enumi}.}
\item
  Easy: FAM ∈ {[}5.5,∞{]}, CNC ∈ {[}5.5,∞{]}. Prioritizes high-frequency words.
\item
  Medium: FAM ∈ {[}4.5,∞{]}, CNC ∈ {[}4.5,∞{]}. Matches the original Perrin et al. (2014) criteria.
\item
  Hard: FAM ∈ {[}3.0,4.5{]}, CNC ∈ {[}3.0,4.5{]}, VAL ∈ {[}2.0,4.0{]}. Strictly isolates lower-frequency, more abstract, and lower-valence words to induce cognitive load.
\end{enumerate}

\section*{5. Extended Capabilities}

While the primary objective is MRS measurement, the architecture of this toolbox allows for parametric manipulation of sentence difficulty to probe specific cognitive deficits.

\subsection*{5.1 Modulating Psycholinguistic Load}

By inverting the standard filters using the band-pass capability, researchers can create "stress tests." Lowering the familiarity threshold recruits rare words (e.g., abacus) to test lexical retrieval. Lowering the concreteness threshold admits abstract concepts (e.g., justice) to test verbal processing systems independent of visual imagery (Paivio, 1971). Furthermore, the valence filter can be inverted to select negative words (e.g., weapon), allowing for the study of affective interference in reading (Egidi \& Gerrig, 2009).

\subsection*{5.2 Manipulating Semantic Distance}

The standard logic determines "False" sentences based on categorical disjointness (e.g., "A dog is a fruit"), which is cognitively effortless to verify. This toolbox extends this logic by allowing manipulation of the semantic distance---the number of edges traversing the ontology tree.

Semantic Interference: By selecting "False" pairs with minimal tree distance, such as siblings sharing a direct parent (e.g., "A dog is a cat"), semantic interference effects similar to the Stroop task can be introduced.

Taxonomic Precision: Conversely, "True" sentences can be manipulated by depth. Verifying "A dog is an animal" (distance = 2) relies on broad category knowledge, whereas "A dog is a mammal" (distance = 1) requires precise taxonomic information.

\section*{6. Validation and Open Science}

The original methodology was validated by Perrin et al. (2014) using a French corpus (N = 41). They demonstrated a strong correlation (r=0.836, p\textless0.001) between the generator\textquotesingle s "True/False" silent reading task and the traditional oral MNREAD test. Furthermore, they observed that the "True/False" verification modality produced results statistically equivalent to oral reading, confirming that this method successfully dissociates reading speed from speech production.

It is important to note that psycholinguistic norms and reading behaviors can vary across languages. While we believe that the methodology transfer is sound, this specific English implementation currently awaits formal psychometric validation against established English reading charts, such as the English MNREAD or IReST (Trauzettel-Klosinski et al., 2012). We provide this toolbox as a complete, open-source package to the scientific community to facilitate this validation process.

\section*{7. Conclusion}

The English implementation of the Perrin Sentence Generator offers a versatile platform for reading assessment. By combining a strict logical ontology with adjustable band-pass psycholinguistic filtering and specific grammatical safeguards for English, it resolves the "finite corpus" problem of traditional charts. Whether used to measure the maximum reading speed of a low-vision patient or manipulated to study the cognitive costs of processing abstract language, this toolbox provides the necessary control and infinite variability required for rigorous experimental design.

\section*{Third-Party Content Notice}

This project incorporates third-party materials governed by their own license terms. The Glasgow Norms dataset (Scott, G. G., Keitel, A., Becirspahic, M., Yao, B., \& Sereno, S. C., 2019) is used under the Creative Commons Attribution 4.0 International License (CC BY 4.0). Copyright © 2018 The Author(s). The original dataset is available at https://doi.org/10.3758/s13428-018-1099-3. Users of this toolbox who redistribute the Glasgow Norms data should retain this license and copyright information and refer to the original source.

\section*{References}

Ahn, S. J., Legge, G. E., \& Luebker, A. (1995). Printed cards for measuring low-vision reading speed. Vision Research, 35(13), 1939--1944.

Baumeister, R. F., Bratslavsky, E., Finkenauer, C., \& Vohs, K. D. (2001). Bad is stronger than good. Review of General Psychology, 5(4), 323--370.

Crossland, M., Legge, G., \& Dakin, S. (2008). The development of an automated sentence generator for the assessment of reading speed. Behavioral and Brain Functions, 4(1), 14.

Egidi, G., \& Gerrig, R. J. (2009). How valence affects language processing: Negativity bias and mood congruence in narrative comprehension. Memory \& Cognition, 37(5), 547--555.

Frijda, N. H. (1986). The emotions. Cambridge University Press.

Gernsbacher, M. A. (1984). Resolving 20 years of inconsistent interactions between lexical familiarity and orthography, concreteness, and polysemy. Journal of Experimental Psychology: General, 113(2), 256--281.

Gilhooly, K. J., \& Logie, R. H. (1980). Age-of-acquisition, imagery, concreteness, familiarity, and ambiguity measures for 1,944 words. Behavior Research Methods \& Instrumentation, 12(4), 395--427.

Kensinger, E. A., \& Corkin, S. (2004). Two routes to emotional memory: Distinct neural processes for valence and arousal. Proceedings of the National Academy of Sciences, 101(9), 3310--3315.

Legge, G. E., Ross, J. A., Luebker, A., \& LaMay, J. M. (1989). Psychophysics of reading. VIII. The Minnesota Low-Vision Reading Test. Optometry and Vision Science, 66(12), 843--853.

Paivio, A. (1971). Imagery and verbal processes. Holt, Rinehart, and Winston.

Paivio, A., Yuille, J. C., \& Madigan, S. A. (1968). Concreteness, imagery, and meaningfulness values for 925 nouns. Journal of Experimental Psychology, 76(1, Pt. 2), 1--25.

Perrin, J.-L., Paillé, D., \& Baccino, T. (2014). A new sentence generator providing material for maximum reading speed measurement. Behavior Research Methods, 46(4), 1036--1048.

Russell, J. A. (1980). A circumplex model of affect. Journal of Personality and Social Psychology, 39(6), 1161--1178.

Scott, G. G., Keitel, A., Becirspahic, M., Yao, B., \& Sereno, S. C. (2019). The Glasgow Norms: Ratings of 5,500 words on nine scales. Behavior Research Methods, 51(3), 1258--1270.

Seiple, W., Szlyk, J. P., McMahon, T., Pulido, J., \& Fishman, G. A. (2005). Eye-movement training for reading in patients with age-related macular degeneration. Investigative Ophthalmology \& Visual Science, 46(8), 2886--2896.

Trauzettel-Klosinski, S., Dietz, K., \& the IReST Study Group. (2012). Standardized assessment of reading performance: The New International Reading Speed Texts IReST. Investigative Ophthalmology \& Visual Science, 53(9), 5452--5461.

\end{document}